\documentclass[preprint,amsmath,amssymb,aps,superscriptaddress]{revtex4-1}

\usepackage{graphicx}% Include figure files
\usepackage{dcolumn}% Align table columns on decimal point
\usepackage{bm}% bold math
\usepackage{color}
\usepackage{multirow}
\usepackage{epstopdf}
\definecolor{greenGM}{rgb}{0.4353, 1, 0}
\definecolor{greenmax1}{rgb}{0.7843,0,0.7843}
\definecolor{greenmax2}{rgb}{0,0.6627,0.4118}
\definecolor{greenenv}{rgb}{0,0.8,0}
\definecolor{lightgray}{rgb}{0.65,0.65,0.65}
\definecolor{lightred}{rgb}{1,0.65,0.65}
\definecolor{red2000}{rgb}{1,0,0.2}
\definecolor{red200}{rgb}{1,0,0.4}
\definecolor{red20}{rgb}{1,0,0.6}
\definecolor{grayL}{rgb}{0.3,0.3,0.3}
\definecolor{grayLmk}{rgb}{0.6,0.6,0.6}
\definecolor{redL}{rgb}{1,0,0.3}
\definecolor{redLmk}{rgb}{1,0,0.6}
\definecolor{boxa}{rgb}{1,0.5,0}
\definecolor{boxs}{rgb}{0.5,0.5,0.5}
\definecolor{boxm}{rgb}{0,0.75,0.2}

\begin{document}

\preprint{APS/123-QED}

\title{Enhancing phonon flow through 1D interfaces by Impedance Matching}

\author{Carlos A. Polanco}
\email{cap3fe@virginia.edu}
\affiliation{Department of Electrical and Computer Engineering, University of Virginia, Charlottesville,VA-22904.}%

\author{Avik W. Ghosh}
\email{ag7rq@virginia.edu}
\affiliation{Department of Electrical and Computer Engineering, University of Virginia, Charlottesville,VA-22904.}%

\date{\today}% It is always \today, today,

\begin{abstract}
We extend concepts from microwave engineering to thermal interfaces and explore the principles of impedance matching in 1D. The extension is based on the generalization of acoustic impedance to non linear dispersions using the contact broadening matrix $\Gamma(\omega)$, extracted from the phonon self energy. For a single junction, we find that for coherent and incoherent phonons the optimal thermal conductance occurs when the matching $\Gamma(\omega)$ equals the Geometric Mean (GM) of the contact broadenings. This criteria favors the transmission of both low and high frequency phonons by requiring that (1) the low frequency acoustic impedance of the junction matches that of the two contacts by minimizing the sum of interfacial resistances; and (2) the cut-off frequency is near the minimum of the two contacts, thereby reducing the spillage of the states into the tunneling regime. For an ultimately scaled single atom/spring junction, the matching criteria transforms to the arithmetic mean for mass and the harmonic mean for spring constant. The matching can be further improved using a composite graded junction with an exponential varying broadening that functions like a broadband antireflection coating. There is however a trade off as the increased length of the interface brings in additional intrinsic sources of scattering.
\end{abstract}

\maketitle

%\tableofcontents

\section{Introduction}
Thermal management of devices at the nanoscale requires a comprehensive understanding of interfacial phonon transport. At length scales on the order of tens of nanometers, phonons flow quasi-ballistically and scattering at interfaces plays a dominant role on the overall thermal conductance \cite{Cahill03,Toberer12,Radfar12,English12,Cahill14}. Many applications stand to benefit from decreasing interfacial scattering; for instance, the heat generated in current transistors is blocked by their low thermal conductance \cite{Haensch06,Pop10},  raising device temperature and impacting performance and reliability \cite{Vasileska08}. One way to  increase interfacial conductance is by inserting a thin junction layer that allows a gradual change of material properties and bridges phonons across the junction. A recent experimental study on Au/Si interfaces showed a 2 fold enlargement on interfacial thermal conductance by the addition of a thin Ti layer ($\approx 7nm$) \cite{Duda13}. Nevertheless, the principles to choose the junction material that maximizes interfacial thermal conductance remain unclear. Finding those principles is not an easy task, since they need to achieve the largest transmission for a { \it broad band spectrum} of phonons, with {\it non-linear dispersion} and well defined translational and rotational {\it symmetries}. 

Some studies have already provided clues on junction properties that correlate with increasing interfacial conductance. It was suggested that maximizing the density of states overlap maximizes thermal conductance \cite{English12,Tian12,Liang12,Zhou13}. Also, it was proposed that maximum conductance requires the junction energy flux to equal the Geometric Mean (GM) of the contact fluxes \cite{Nam12}. For atomic junctions, the maximum happens for the Arithmetic Mean (AM) of the contact masses and the Harmonic Mean (HM) of the contact spring constants \cite{Saltonstall13,Zhang11}. In spite of the many advances towards finding the junction material that maximizes interfacial thermal conductance, a unifying and quantitative physical picture is still missing. 

\begin{figure}[htb]
	\centering
	\includegraphics[width=86mm]{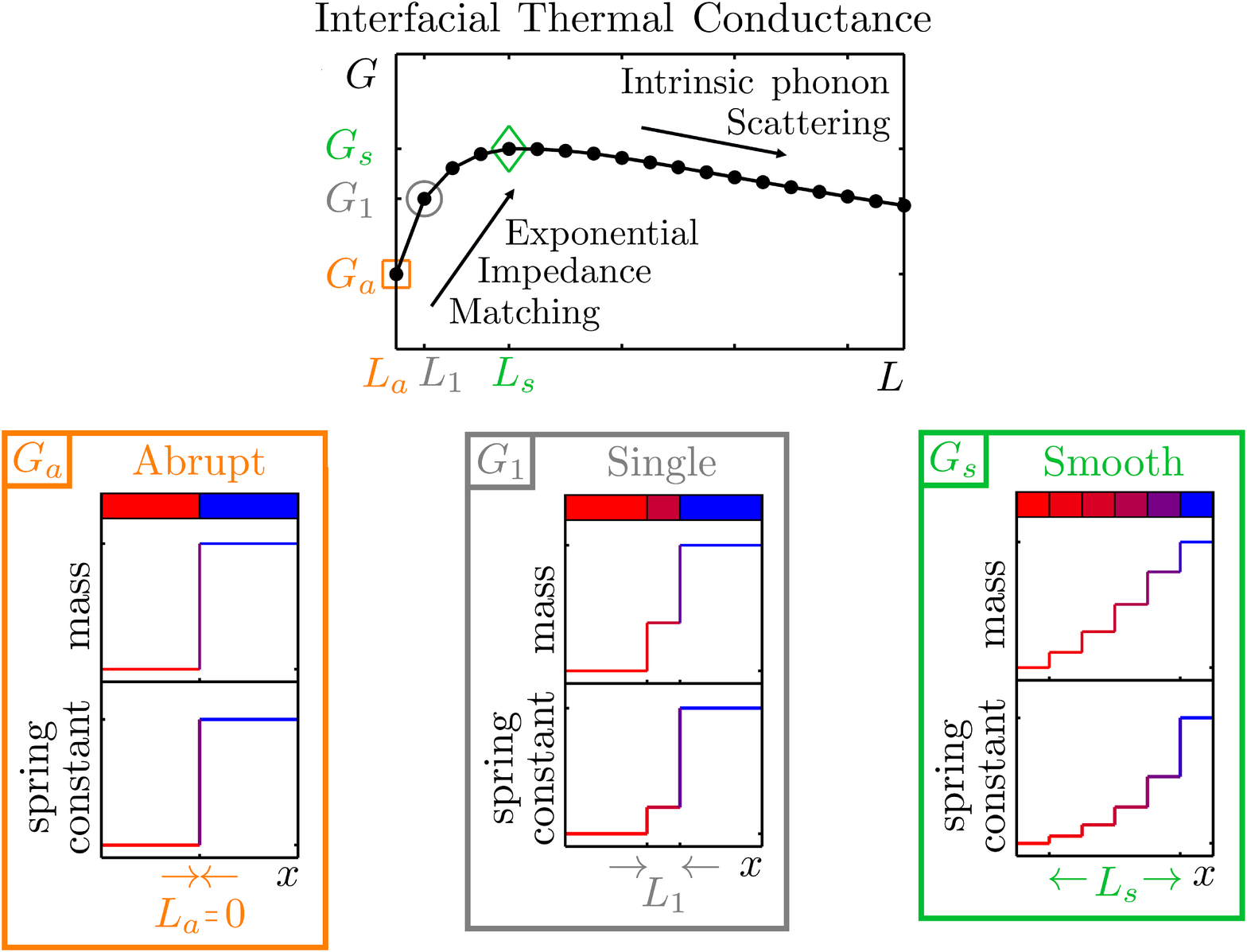}
	\caption{Interfacial thermal conductance is maximized by inserting a graded junction with an exponentially varying ``impedance'' $\Gamma_n=\Gamma_le^{\zeta n}$ (Eq.~\ref{equevGamma}). However as the length between the contacts increases, the conductance gained by ``impedance'' matching is dominated by losses due to other scattering mechanisms.}
	\label{figintro}
\end{figure} 

In this paper we explain how to choose the mass, spring constant and length $(m,k,L)$ of a single or several junction materials to maximize the average of phonon transmission over frequency $\left\langle T\right\rangle_\omega$, which is equivalent to maximizing the thermal conductance of 1D chains (Fig.~\ref{figintro}). Our analysis extends the acoustic impedance to the non-linear dispersion limit using the contact-induced broadening matrix $\Gamma$, obtained from the imaginary part of the self-energy matrix in the Non-Equilibrium Green's Function (NEGF) formalism \cite{Polanco13}. 

For a single junction material, we find that the desirable broadening is the Geometric Mean (GM) of the contact broadenings. In the coherent limit, this choice maximizes the resonance peaks of the transmission function (Sec.~\ref{sec1Mcoh}), while in the incoherent limit it minimizes the sum of interfacial resistances (Sec.~\ref{sec1Mincoh}). To satisfy the GM condition we need (1) to choose the acoustic impedance as the GM of the contact impedances to favor low frequency phonons and (2) to set the cut-off frequency close to the minimum of the contact cut-off frequencies to favor high frequecy phonon and reduce the spillage of states into the tunelling regime (Fig.~\ref{figZvsGamma}a). In Sec.~\ref{sec1Mcoh}, those sub conditions are translated into rules to choose the junction mass, spring constant, group velocity, Debye temperature or density of states. As the junction becomes a single atom or bond, the resonances are pushed outside the allowed frequency interval and the GM condition transforms to the Arithmetic Mean (AM) of the masses or the Harmonic Mean (HM) of the spring constants of the contacts. 

For multi-junction materials, the GM condition evolves into an exponentially varying broadening, since the GM of two quantities stays near the lower partner. That variation generates a {\it broadband anti-reflection coating} in the coherent limit (Fig.~\ref{figTNM}b Sec.~\ref{secNMcoh}) and minimizes the sum of resistances between dissimilar materials in the incoherent limit (Sec.~\ref{secNMincoh}). In this way, thermal conductance is pushed to the maximum physical limit allowed by the contacts as the number of junctions increases. Finally, we discuss the trade-off between increasing conductance by adding junctions to improve the matching of the contacts and decreasing conductance as a result of enlarging the interface and adding intrinsic scattering (Fig.~\ref{figintro}).

The results of this paper can be used for modeling thermal conductance of Self-Assembled Monolayers (SAMs) and for extending 1D propagation models of long-wavelength phonons in semiconductors \cite{Leon01,Villegas08}. It also provides a solid ground to expolore the 3D maximization problem since a 3D system can be decomposed into decoupled ``1D like" subsystems, with the caveat that in each of them the nature of the polarization vectors and their overlap should play a role, depending on the symmetry breaking disorder at the interface. The exponentially varying broadening provides one example of phonon transmission engineering inspired by microwave engineering of broad band filters \cite{Pozar04}. Similar analogies can be found for the binomial or Chebyshev matching transformers, which optimize the flatness or bandwidth of the transmission function respectively \cite{Pozar04}. We can further envision engineering an ultra narrow transmission that filters particular phonons or an extremely low transmission to improve thermoelectric figure of merit.

\section{Methodology}

Thermal conductance $G^Q$ is defined as the ratio between heat current $I^Q$ and temperature difference across the contacts $\Delta {\bf T}$, which according to Landauer formalism is given by \cite{Jeong12}
\begin{equation}
G^Q=\int_0^\infty \frac{d\omega}{2\pi}\hbar\omega MT \frac{\partial N}{\partial {\bf T}},
\label{equGQ}
\end{equation}
where $\hbar\omega$ is the energy of a phonon, $M$ is the number of conducting channels, $T$ is the average transmission per channel and $N$ the Bose-Einstein distribution. The factor $MT$ can be found using NEGF formalism from  $MT=\text{Trace}\left\{\Gamma_1G\Gamma_2G^\dagger\right\}$, with $G$ the retarded Green's function of the system of interest and $\Gamma_i$ the broadening matrix, which is the anti-hermitian part of the self-energy and describes the interaction of the system with contact $i$ \cite{Mingo03,Datta05}. The number of conducting channels depends on the dimension and force constants of the system. In particular, for a 1D system with 1st neighbor interactions $M$ reduces to one. We further simplify Eq.~\ref{equGQ} by assuming a temperature larger than the Debye temperature, so the factor $\hbar\omega \partial N/\partial {\bf T}$ is almost constant over the allowed frequency spectrum. In this way, $G^Q\propto \langle T \rangle_{\omega}$ and maximizing  $\langle T \rangle_{\omega}$ maximizes $G^Q$ (Table~\ref{tab1}).

In spite of our second assumption  to simplify $G^Q$, our results are not limited to high temperatures. As we explain in Sec.~\ref{sec1Mcoh} and \ref{sec1Mincoh},  the necessity of finding a robust maximum requires us to maximize $\langle \langle T \rangle_\varphi\rangle_{\omega}$ the frequency average of the phase average of $T$. The phase average transforms $T$ into a slow varying function by washing away the interference wiggles. Therefore, for the maximization process we can pull $\langle T \rangle_\varphi$ the phase average of $T$ out of the integral and Eq.~\ref{equGQ} can be approximated by
\begin{equation}
G^Q\approx\langle \langle T \rangle_\varphi\rangle_{\omega}\int\limits_0^{\omega_{max}} \frac{d\omega}{2\pi}\hbar\omega \frac{\partial N}{\partial {\bf T}}.
\label{equapprox}
\end{equation}
Now that the temperature dependence has been isolated, maximizing $\langle \langle T \rangle_\varphi\rangle_{\omega}$ yields to the more robust maximum of $G^Q$. The approximation in Eq.~\ref{equapprox} breaks down for very low temperatures because the frequency average should be done only over the exited phonons which constitutes a frequency range shorter than $[0,\omega_{max}]$. Nevertheless, the material properties that maximize $\langle \langle T \rangle_\varphi\rangle_{\omega}$ only differ from the exact solution for temperatures below 10\% of the Debye temperature.

\begin{table}[hbt]
	\begin{tabular}{|c|c|}
		\hline
		$T$ & Coherent transmission \\ \hline
		$\langle T\rangle_{\omega}$ & Frequency average of $T$ over $[0,\omega_{max}]$ \\ \hline
		$\langle T\rangle_{\varphi}$ & Incoherent transmission or phase average of $T$ \\ \hline
		$\langle \langle T\rangle_\varphi \rangle_{\omega}$ & Frequency average of the phase average of $T$ \\ \hline
	\end{tabular}
	\caption{Summery of the notation for the different types of phonon transmission used in the document. The maximum freuquency $\omega_{max}$ is defined by the minimum cut-off frequency of the contacts.}
	\label{tab1}
\end{table}

\section{Single material, coherent.\\ Geometric Mean and anti-reflection coating.} \label{sec1Mcoh}

\begin{figure}[htb]
	\centering
	\includegraphics[width=86mm]{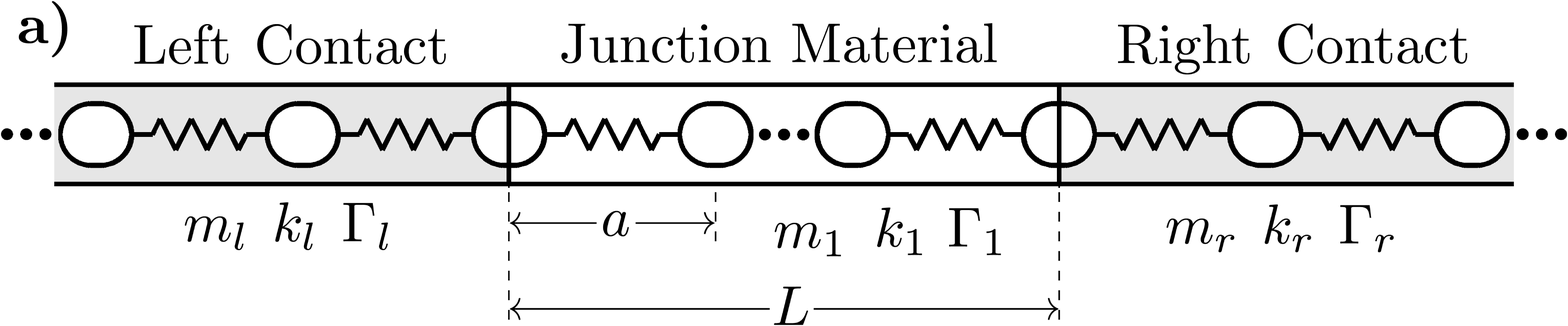}
	\includegraphics[width=86mm]{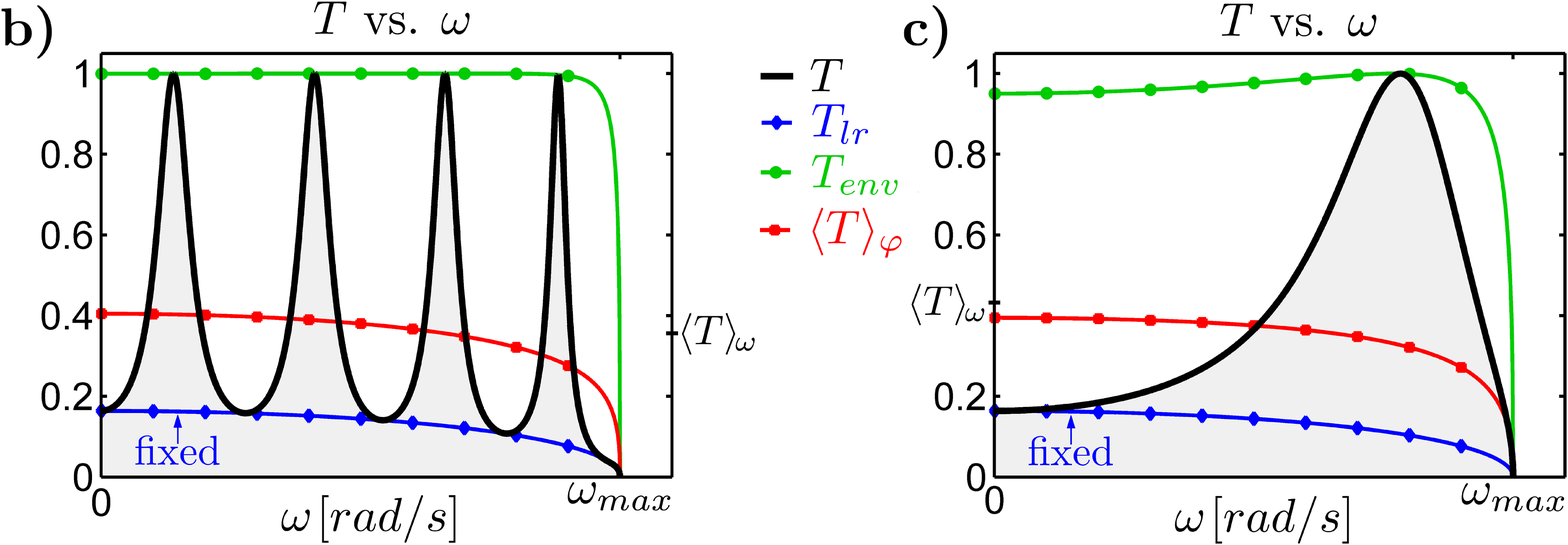}
	\caption{{\bf a)} 1D crystal modeled as a chain of masses joined by springs. {\bf b)} and {\bf c)} The choice that maximizes $\left\langle T \right\rangle_{\omega}$ is an interplay between pushing the envelope up and fitting the right number of resonances inside the allowed frequency range. The figure shows an example where the largest envelope generated by the GM of contact broadenings {\bf (b)} does not guarantee maximum $\left\langle T \right\rangle_{\omega}$ {\bf (c)}. It also shows the transmission for the abrupt interface between contacts $T_{lr}$ and the incoherent transmission $\langle T \rangle_{\varphi}$, which washes away interference (no wiggles). }
	\label{figT3M}
\end{figure}

Analogous to the quantum mechanical transmission of a particle through a barrier, the phonon transmission function $T$ for the system in Fig.~\ref{figT3M}a is found from the ratio between the incident and transmitted heat currents. After assuming plane waves solutions, imposing boundary conditions at the interfaces and replacing the acoustic impedance $Z$ with the broadening matrix $\Gamma$ \cite{Polanco13}, $T$ is given by 
\begin{equation}	
T=\frac{4\gamma_{lr}}{(\gamma_{lr}+1)^2\cos^2(q_1L)+(\gamma_{l1}+\gamma_{1r})^2\sin^2(q_1L)},
\label{equTph3M}
\end{equation}
where the wave vector of the junction $q_1$ comes from the dispersion relation, i.e. $\omega=\omega_{c1}\left|\sin(q_1a/2)\right|$ with the cut-off frequency $\omega_{c1}=2\sqrt{k_1/m_1}$, and $\gamma_{\alpha\beta}=\Gamma_\alpha/\Gamma_\beta$. The broadening function $\Gamma$ for a 1D chain \cite{Tsegaye} describes how easily phonons can move from one site to the next and is related with the acoustic impedance through \cite{Polanco13}
\begin{equation}
\Gamma=2\omega\rho v_g(\omega)=2\omega Z(\omega),
\label{equGamma}
\end{equation}
with $\rho$ the mass density and $v_g$ the group velocity. Eq.~\ref{equTph3M} is valid for frequencies smaller than all the cut-off frequencies. When $\omega_{c1}<\omega<\omega_{cl},\omega_{cr}$, phonons tunnel through the junction and the $\cos$ and $\sin$ become their hyperbolic counterparts. When the frequency is larger than the maximum frequency $\omega>\min(\omega_{cl},\omega_{cr})=\omega_{max}$, $T=0$ defining the allowed frequency range. 

Note that the transmission function $T$ also follows from adding coherently all the wave paths that cross the interface
\begin{align}
T=\gamma_{rl}\left|t\right|^2=&\gamma_{rl}\left|t_{l1}t_{1r}+t_{l1}r_{1r}r_{1l}t_{1r}e^{i2q_1L}+\cdots \right|^2 \\
=& \frac{T_{l1}T_{1r}}{1+2\sqrt{R_{1r}R_{1l}}\cos(2q_1L)+R_{1r}R_{1l}} \nonumber,
\end{align}
where at each abrupt interface, the reflection and transmission coefficients are $r_{\alpha\beta}=(\gamma_{\alpha\beta}-1)/(\gamma_{\alpha\beta}+1)$ and $t_{\alpha\beta}=(2\gamma_{\alpha\beta})/(\gamma_{\alpha\beta}+1)$, and the reflection and transmission are $R_{\alpha\beta}=|r_{\alpha\beta}|^2$ and $T_{\alpha\beta}=\gamma_{\beta\alpha}|t_{\alpha\beta}|^2$.

As shown in Fig.~\ref{figT3M}b, $T$ is an oscillating function bounded by the transmission of the abrupt interface between contacts 
\begin{equation}
T_{lr}=\frac{4\gamma_{lr}}{(\gamma_{lr}+1)^2},
\end{equation}
which is prespecified, and an envelope function that depends on our choice of junction 
\begin{equation}
T_{env}=\frac{4\gamma_{lr}}{(\gamma_{l1}+\gamma_{1r})^2}.
\end{equation}
Therefore, as long as $\Gamma_1$  lies in between $\Gamma_l$ and $\Gamma_r$, $T_{env}>T_{lr}$, $\left\langle T \right\rangle_\omega>\left\langle T_{lr} \right\rangle_\omega$ and the ballistic thermal conductance of the gradual interface is larger than the abrupt interface. 

Similar to an anti-reflection coating, we can eliminate reflection  {\it at a single frequency} by destructive interference of the wave path undergoing a single reflection with the sum of paths undergoing more than one. The reflection coefficient $r$ is given by the sum of all the wave paths that return to the incident contact
\begin{align}
r&=r_{l1}+t_{l1}r_{1r}t_{1l}e^{i2q_1L}+t_{l1}r^2_{1r}r_{1l}t_{1l}e^{i4q_1L}+\cdots \nonumber \\
&=r_{l1}+r_{1r}\frac{t_{l1}t_{1l}}{1-r_{1r}r_{1l}e^{i2q_1L}}. \label{equrwavepath}
\end{align}
To eliminate reflection $R=|r|^2=0$, the factors in Eq.~\ref{equrwavepath} must have equal magnitude and opposite phase. The first condition requires the junction broadening to be the GM of the contact broadenings $\Gamma_1=\sqrt{\Gamma_l\Gamma_r}$, and the second condition happens if the length of the junction $L_1$ fits exactly a quarter wavelength.

The choice that maximizes $\left\langle T \right\rangle_{\omega}$ is an interplay between pushing $T_{env}$ up  and fitting the right number of resonances, since our choice of $(m_1,k_1,L_1)$ affects both the envelope function $T_{env}$ and the position of the resonances.  Intuitively one might think that maximizing the envelope {\it over the allowed frequency interval}, by making $\Gamma_1$ the GM of the contact broadenings, generates the largest $\left\langle T \right\rangle_{\omega}$. However on a finite frequency range, the average fluctuates according to the number of resonances within the interval, with larger fluctuations for fewer resonances. For instance, the maximum $\left\langle T \right\rangle_{\omega}$ happens for a single resonance around $\omega_{max}/(3/2-2/3\pi^2)$ (Fig.~\ref{figT3M}c). This maximum is not very robust since it relies on our ability to place a single resonance at a particular frequency, which is limited by our experimental accuracy on $m_1,k_1$ and $L_1$.

A more robust condition consists of choosing $\Gamma_1$ as the GM of the contact broadenings {\it over the allowed frequency interval}. This condition maximizes the integral over $\omega$ of the phase $(\varphi=q_1L)$ average of the transmission $\langle T\rangle_\varphi$ (See Eq.~\ref{equTphase}) 
\begin{equation}
\Gamma_{GM}(m_{GM},k_{GM})\leftrightarrow \max_{\Gamma_1(m_1,k_1)}  \int\limits_0^{\omega_{max}}d\omega \sqrt{T_{lr}}\sqrt{T_{env}},
\label{equGMcondition}
\end{equation}
which is equivalent to pushing the envelope function up (Fig.~\ref{figT3M}b). Although choosing the GM condition does not guarantee the maximum $\left\langle T\right\rangle_\omega$, its difference with the maximum is bounded by $\left\langle T\right\rangle_{max}-\left\langle T_{GM}\right\rangle_\omega \lesssim 0.1(1-\left\langle T_{GM}\right\rangle_\omega)$. But more importantly, the condition is more robust because it does not rely on interference effects. 

\begin{figure}[htb]
	\centering
	\includegraphics[width=86mm]{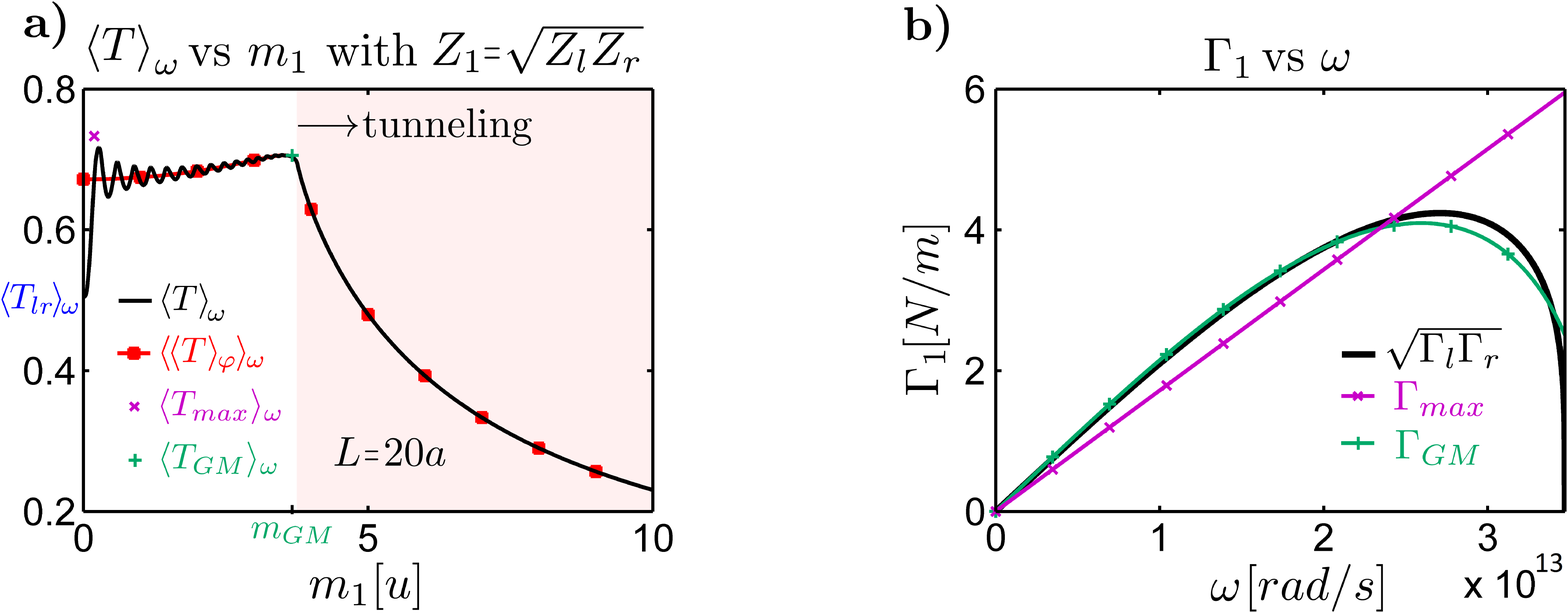}
	\caption{{\bf (a)} Choosing the impedance $Z_1=\sqrt{Z_lZ_r}$ does not guarantee high $\left\langle T\right\rangle_\omega$ because if the cut-off frequency lies below $\omega_{max}$, tunneling cuts down transmission (shaded region). On the other hand, choosing the broadening $\Gamma_1\approx\sqrt{\Gamma_l\Gamma_r}$ {\bf (b)} fixes a unique pair $m_{GM},k_{GM}$ that favors the flow of both low and high frequency phonons. The figure shows a similar result for the incoherent transmission average $\langle\langle T\rangle_\varphi\rangle_\omega$. Note that $\Gamma$ for the maximum $\left\langle T\right\rangle_\omega$ ($\Gamma_{max}$) intersects $\sqrt{\Gamma_l\Gamma_r}$ where the single resonance is located (Fig.~\ref{figT3M}c).}
	\label{figZvsGamma}
\end{figure}

{\it The broadening for the GM condition $\Gamma_{GM}$ favors the flow of both low and high frequency phonons} by making $\Gamma_{1}$  close to $\sqrt{\Gamma_l\Gamma_r}$ over the entire frequency range (Fig.~\ref{figZvsGamma}b). Note that the GM of acoustic impedance by itself does not guarantee high $\left\langle T\right\rangle_\omega$ (Fig.~\ref{figZvsGamma}a). Indeed, if our choice of $m_1,k_1$ makes $\omega_{c1}<\omega_{max}$, phonon transmission is drastically cut by tunneling.

A back of the envelope estimation for $m_{GM}$ and $k_{GM}$ follows by graphically overlapping $\Gamma_1(m_1,k_1)$ with $\sqrt{\Gamma_l\Gamma_r}$ (Fig.~\ref{figZvsGamma}b). For low frequencies the slope of $\Gamma$ is dictated by the impedance, so to favor the flow of low frequency phonons we want $Z_1=\sqrt{Z_lZ_r}$. To favor the flow of high frequency phonons, we equate the frequency $\omega^*$ at which $\sqrt{\Gamma_l\Gamma_r}$ is maximum to the frequency at which $\Gamma$ is maximum $\omega_{c1}/\sqrt{2}=\omega^*$. Then, from $Z=\sqrt{mk}$ and $\omega_c=2\sqrt{k/m}$ we solve for $m_{GM}$ and $k_{GM}$ 
\begin{equation}
k_{GM}\approx\frac{\omega_{c1}Z_1}{2}\ \ \ \ \ \ m_{GM}\approx\frac{2Z_1}{\omega_{c1}}.
\label{equbea1}
\end{equation}
From Eq.~\ref{equbea1} we can also infer how to choose the junction group velocity $v_g=a\sqrt{k/m}$ and the Debye temperature $T_D=\hbar\omega_D/k_B$ with $\omega_D=\omega_c\pi/2$ and $k_B$ the Boltzmann constant.  Using Eq.~\ref{equGamma}, we can also express the GM condition in terms of the density of states per unit cell ($DOS$)
\begin{equation}
\Gamma=\frac{2\omega m}{\pi DOS},
\label{equGammaDOS}
\end{equation}
which quantifies how to choose the overlap of density of states to maximize thermal conductance. 

When the junction material becomes a single mass or bond, the GM condition to maximize $\left\langle T\right\rangle_\omega$ transits to the AM of the contact masses or the HM of the contact spring constants respectively. This transition happens when $k_1$ increases enough or $m_1$ decreases enough to push $T_{env}$ below $T_{lr}$ (Fig.~\ref{figT3M}) making $\left\langle T\right\rangle_\omega<\left\langle T_{lr}\right\rangle_\omega$. Then it is better to have an abrupt interface between the contacts which take us back to the AM or HM condition. In particular, if the spring constant $k_1>>\omega_{max}^2m_1(L/a)^2$, then $\sin^2(q_1L)\approx \omega^2 m_1 L^2/k_1a$ and $\Gamma_1\approx 2\omega\sqrt{k_1 m_1}$. Replacing the approximations into Eq.~\ref{equTph3M} and letting $k_1$ tend to infinity we recover the transmission for a single atomic junction \cite{Polanco13}
\begin{equation}
T\approx \frac{4\gamma_{lr}}{(\gamma_{lr}+1)^2+\frac{4\omega^4 m_1^2 (L/a)^2}{\Gamma_r^2}}.
\label{equsmjunc}
\end{equation}
The maximum of Eq.~\ref{equsmjunc} happens when $m_1=0$, which recovers the AM condition for the atomic mass linking the contacts (Fig. \ref{figT3M}a). Another way to rationalize the transition from the GM condition is by realizing that increasing $k_1$ or decreasing $m_1$ pushes the resonances out of the allowed frequency interval. Then, without resonances available to reach $T_{env}$, maximizing $T_{env}$ with the GM condition is no longer useful.

\section{Single material, Incoherent. \\ Still Geometric Mean.} \label{sec1Mincoh}

When the phase gained by a phonon in its transit between interfaces is randomized, interference disappears and we can think of phonons as classical particles. The phase ($\varphi=q_1L$) average of the transmission in Eq.~\ref{equTph3M} $\left\langle T\right\rangle_\varphi$ or the incoherent transmission is given by (Fig.~\ref{figT3M})
\begin{equation}
\left\langle T\right\rangle_\varphi=\frac{4\gamma_{lr}}{(\gamma_{lr}+1)(\gamma_{l1}+\gamma_{1r})}=\sqrt{T_{lr}}\sqrt{T_{env}}.
\label{equTphase}
\end{equation}
Since $T_{lr}$ is prespecified, this expression is once again maximized by the GM condition, for which $T_{env}\approx 1$ and $\langle T_{GM}\rangle_\varphi\approx\sqrt{T_{lr}}>T_{lr}$ (Fig.~\ref{figT3M}b). Note that Eq.~\ref{equTphase} is the same expression obtained by adding incoherently all the possible wave paths from the left to the right contact
\begin{equation}
\left\langle T\right\rangle_\varphi=T_{l1}T_{1r}+T_{l1}R_{1r}R_{1l}T_{1r}+\cdots=\frac{T_{l1}T_{1r}}{1-R_{1l}R_{1r}}
\label{equknown}
\end{equation}
with $T_{\alpha\beta}=4\gamma_{\alpha\beta}/(\gamma_{\alpha\beta}+1)^2$, the transmission at the interface between material $\alpha$ and $\beta$, and $R_{\alpha\beta}=1-T_{\alpha\beta}$ the reflection.

The GM condition can be rationalized as a minimization of the sum of two opposite resistances, which are proportional to the ratio between phase average reflection and phase average transmission \cite{Anderson80}. Reorganizing Eq.~\ref{equknown} to expose the additive property of resistance \cite{Datta97}
\begin{equation}
\frac{\left\langle R\right\rangle_\varphi}{\left\langle T\right\rangle_\varphi}=\frac{1-\left\langle T\right\rangle_\varphi}{\left\langle T\right\rangle_\varphi}=\frac{1-T_{l1}}{T_{l1}}+\frac{1-T_{1r}}{T_{1r}},
\label{equaddprop}
\end{equation}
it follows that a variation of the junction produces opposite trends on the single interface resistances. Therefore, Eq.~\ref{equaddprop} is minimum when the individual resistances are equal, which leads to the GM condition. 

Phase randomization can arise from lack of experimental control. For instance, the measured thermal conductance on a SAM junction is the sum of transmissions over molecules with slightly different lengths determined by the cross-sectional variation of the SAM. According to the variation strength, the average transmission can be approximated by 
\begin{equation}
\left\langle T\right\rangle_{\varphi_0-\delta}^{\varphi_0+\delta}\approx \left[T(\varphi_0)-\left\langle T\right\rangle_\varphi\right]\frac{\sin(2\delta)}{2\delta}+\left\langle T\right\rangle_\varphi,
\end{equation}
with $2\delta$ a uniformly distributed phase interval around $\varphi_0$ over which we average $T$. This expression follows from finding the effect of phase randomization on the peak of a sinusoidal function and using it to interpolate the two known limits. Note that $\delta=0$ recovers the fully coherent limit Eq.~\ref{equTph3M} while $\delta=\pi/2$ the fully phase randomized limit Eq.~\ref{equTphase}.

In contrast to the coherent limit, the incoherent limit does not provide a condition for the junction length. To be consistent with our assumption that the phase is randomized between interfaces, we need to guarantee that $L_1$ is longer than the phonon phase coherent length. In case the phase randomization happens at the interfaces, $L_1$ is limited by the longest distance between interacting unit cells. That is the smallest block of material that allows us to define impedance.

\section{N Materials, Coherent. \\ Exponential.}\label{secNMcoh}

\begin{figure}[htb]
	\centering
	\includegraphics[width=86mm]{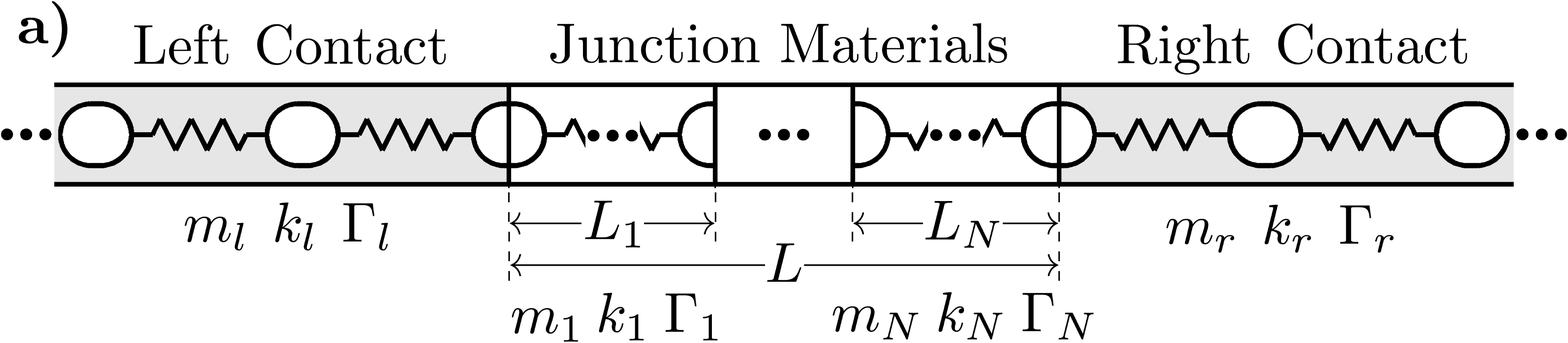}
	\includegraphics[width=86mm]{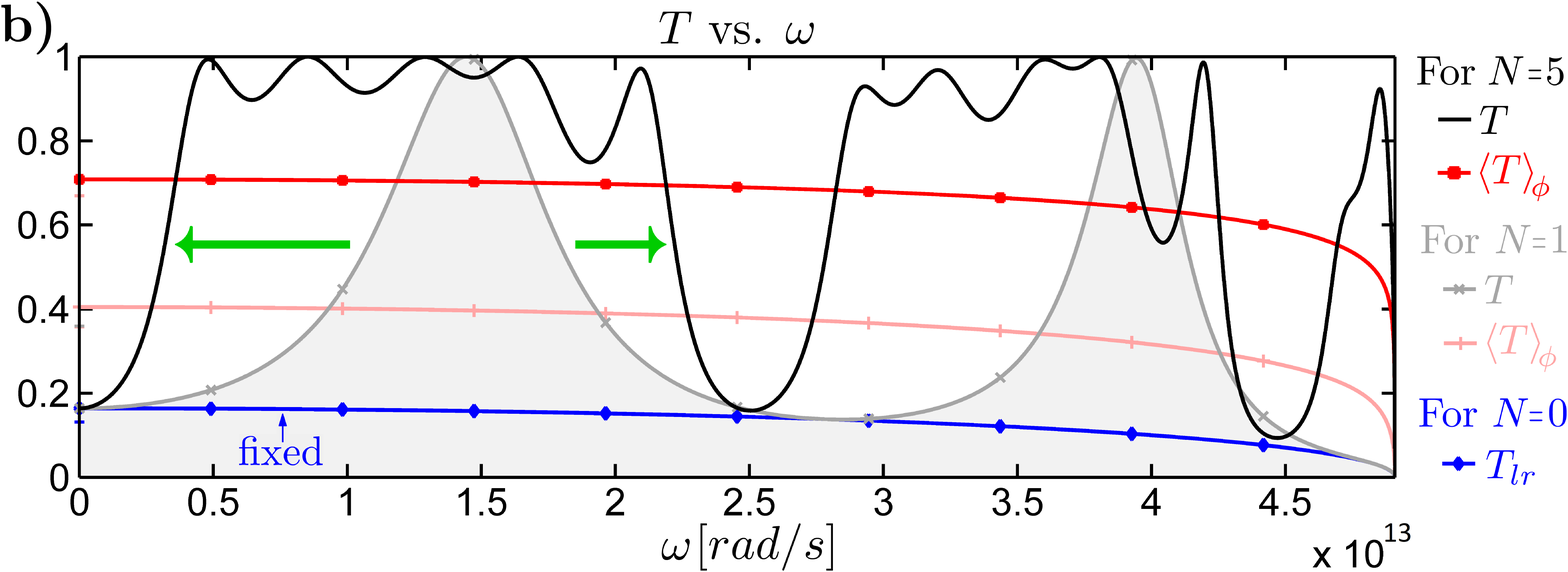}
	\includegraphics[width=86mm]{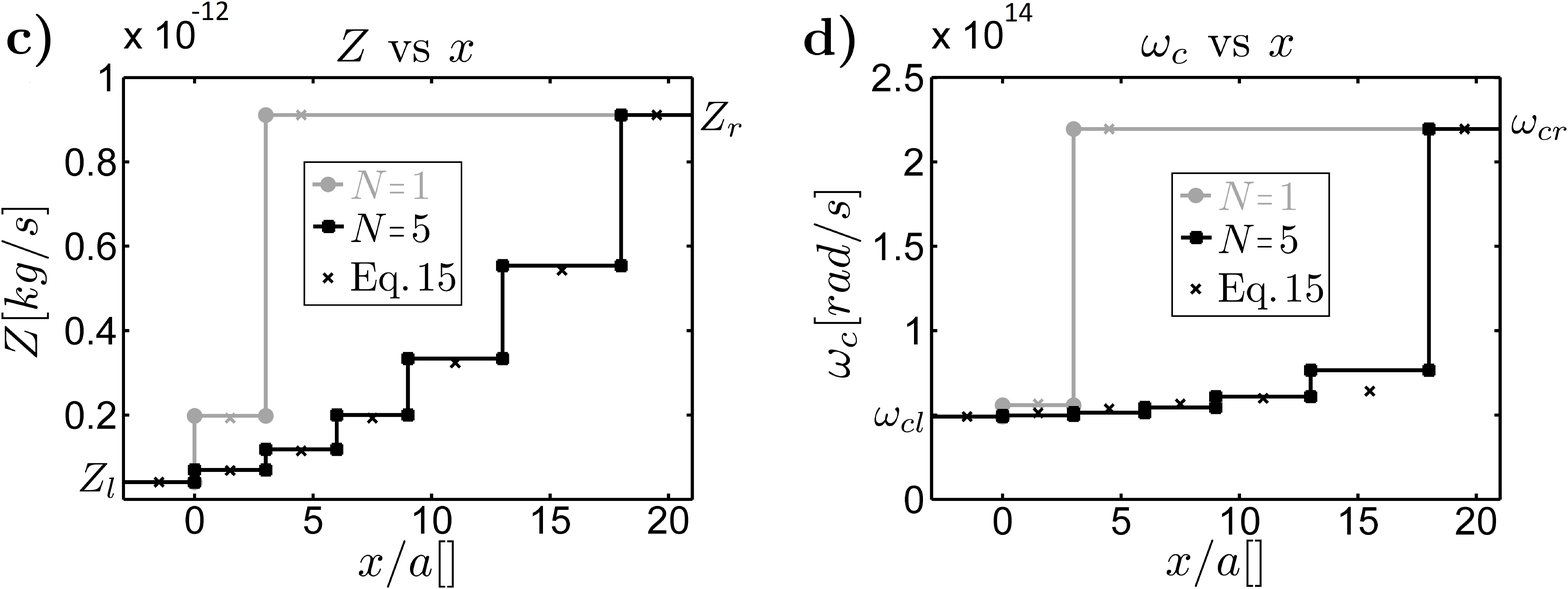}
	\caption{{\bf a)} 1D crystal modeled as a chain of masses joined by springs. {\bf b)} Transmission $T$ and its phase average $\left\langle T\right\rangle_\varphi$ for contacts coupled through an exponential variation of broadenings with a single junction $N=1$ and with $N=5$ junction materials. As we increase the number of materials transmission widens increasing $\left\langle T\right\rangle_\omega$. The spatial variation of acoustic impedance is shown in {\bf c)} and of cut-off frequency in {\bf d)} for $N=1$ and $N=5$. The crosses show the back of the envelope calculation given by Eq.~\ref{equbecNm}.}
	\label{figTNM}
\end{figure}

For several junction materials in between the contacts (Fig.~\ref{figTNM}a), enforcing the GM condition for every three consecutive pieces translates into an exponentially varying broadening (Eq.~\ref{equevGamma}). This trend follows from taylor expanding $\Gamma(x)=\sqrt{\Gamma(x+\Delta x)\Gamma(x-\Delta x)}$ to second order and solving the corresponding differential equation $(\Gamma')^2=\Gamma\Gamma''$. As we increase the number of materials, this variation generates a banded transmission that monotonically widens. In this way, we approach the maximum limit of thermal conductance imposed by the contacts. The idea was inspired by the exponentially tapered impedance coupling used to design broad band transmission lines \cite{Pozar04}, which constitutes just one example of the extrapolation possibilities from Microwave engineering to phonon engineering.

The transmission for a system with $N$ junction materials is a complicated function dictated by interference patterns. However, to maximize transmission, we intuitively expect a small reflection at each interface so we can approximate the reflection coefficient $r$ by the sum of wave paths undergoing a single reflection, i.e.
\begin{equation}
r\approx r_{l1}+r_{12}e^{i2\varphi_1}+r_{23}e^{i2\varphi_2}+\cdots +r_{Nr}e^{i2\varphi_N}.
\label{equr0}
\end{equation}
At this point we have $N$ exponential basis vectors to fit a desired reflection function, which is a rich problem with different answers according to the design criteria for the function. We further simplify the expression by using our intuition of anti-reflection coating, where to eliminate the first reflection $r_{l1}$ we need  $r-r_{l1}$ to have the same magnitude and opposite phase. Then we assume equal phase gained between interfaces, i.e. $\varphi_n=2n\varphi$, to get complex roots of 1 equally spaced in Eq.~\ref{equr0}, and equal reflection at each interface $r_{i,i+1}=r_{i+1,i+2}=r_i$ to assure equal magnitude. Then Eq.~\ref{equr0} turns to
\begin{equation}
r\approx r_{i}\sum_{j=0}^{N}e^{i2j\varphi}=r_i\frac{1-e^{i2(N+1)\varphi}}{1-e^{i2\varphi}}.
\label{equr}
\end{equation}
As $N$ increases, the resonances ($r=0\leftrightarrow e^{i2(N+1)\varphi}=1$), the anti-resonances ($e^{i2(N+1)\varphi}= -1$) and the decreasing trend of $r_i$ generate a reflection that broadens and tends to zero or equivalently a transmission that broadens and tends to unity. Fig.~\ref{figTNM}b shows the widening of the transmission when we increase the number of junctions from $N=1$ to $N=5$.

Equal reflection at each interface follows from imposing the GM condition on every three consecutive materials, which leads to an exponential variation
\begin{equation}
\Gamma_n=\Gamma_l^{\frac{N+1-n}{N+1}}\Gamma_r^{\frac{n}{N+1}}=\Gamma_le^{\zeta n},
\label{equevGamma}
\end{equation}
with $\zeta=\ln(\Gamma_l/\Gamma_r)/(N+1)$. This variation maximizes the frequency average of the incoherent transmission (See Eq.~\ref{equaddprop2} and below). Therefore we find the $m_n$ and $k_n$ that better satisfy Eq.~\ref{equevGamma} by maximizing $\langle\langle T \rangle_{\varphi}\rangle_w$. Similar to Sec.~\ref{sec1Mcoh}, a useful back of the envelope approximation follows by graphically overlapping the desired and possible broadenings. To match the low frequency slope we need $Z_n=Z_l^{N+1-n/N+1}Z_r^{n/N+1}$ and to match the high frequency spectrum we need $\omega_{cn}=\sqrt{2}\omega_n^*$, with $\omega_n^*$ the frequency at the maximum value of Eq.~\ref{equevGamma}. Then $m_n$ and $k_n$ are approximated by
\begin{equation}
k_n\approx\frac{\omega_{cn}Z_n}{2}\ \ \ \ \ \ m_n\approx\frac{2Z_n}{\omega_{cn}}
\label{equbecNm}
\end{equation}
and rules for the variation of the group velocity, Debye temperature and overlap of density of states can be found following Sec.~\ref{sec1Mcoh}. Fig.~\ref{figTNM}c,d show the spatial variation of junction impedances and cut-off frequencies for $N=1$ and $N=5$. {\it Note that the trend of impedance follows an exponential variation that favors flow of low frequency phonons while the trend of cut-off frequencies is stable around $\omega_{max}$ to favor flow of high frequency phonons}.

Equal phase gain between consecutive interfaces needs
\begin{equation}
q_nL_n=\frac{L_n}{a}2\sin^{-1}\left(\frac{\omega}{\omega_{cn}}\right)=\varphi,
\label{equln}
\end{equation}
which we can only satisfy at a single frequency $\omega^{**}$. A careless choice of $\omega^{**}$ and $\varphi$ might lead to impractically small lengths ($L_n\sim 2a$) or long lengths where phonons are not coherent  ($L_n\sim$ mean free path). A sensible choice is for example $\omega^{**}=\omega_{max}/2$ and $\varphi=2\pi$, which generates lengths on the order of $L_n\sim 10a\sim 5nm$. 

The theory of small reflections shown in this section is valid as long as we can neglect the contributions from paths that involve more that one reflection, i.e. $|r_i|^3<<|r_i|$. For $\Gamma_l<\Gamma_r$ and if we consider $|r_i|<0.1$ small enough, the minimum number of junctions $N_{min}$ needed to be in this regime is
\begin{equation}
N_{min}\approx 1.5\ln\left(\frac{\Gamma_r}{\Gamma_l}\right)-1.
\end{equation}
For instance, for a 10-fold mismatch $\Gamma_r/\Gamma_l=10$ $N_{min}\approx2.5$ and for a 100-fold mismatch $N_{min}\approx6$.

\section{N Materials, Incoherent. \\ Back to Exponential.}\label{secNMincoh}

When phonon  phase is randomized in between interfaces, the exponentially varying broadening minimizes resistance and generates maximum $\langle\langle T \rangle_\varphi\rangle_\omega$. Total resistance decreases as the number of junction materials $N$ increases because the addition of resistance coming from new interfaces is dominated by the decrease on interfacial resistance due to less mismatch. Nevertheless, as we increase $N$, the length between contacts increases and other scattering mechanism become important. Therefore, there is a sweet spot for $N$ at which we stop increasing thermal conductance by adding junction materials.

\begin{figure}[htb]
	\centering
	\includegraphics[width=86mm]{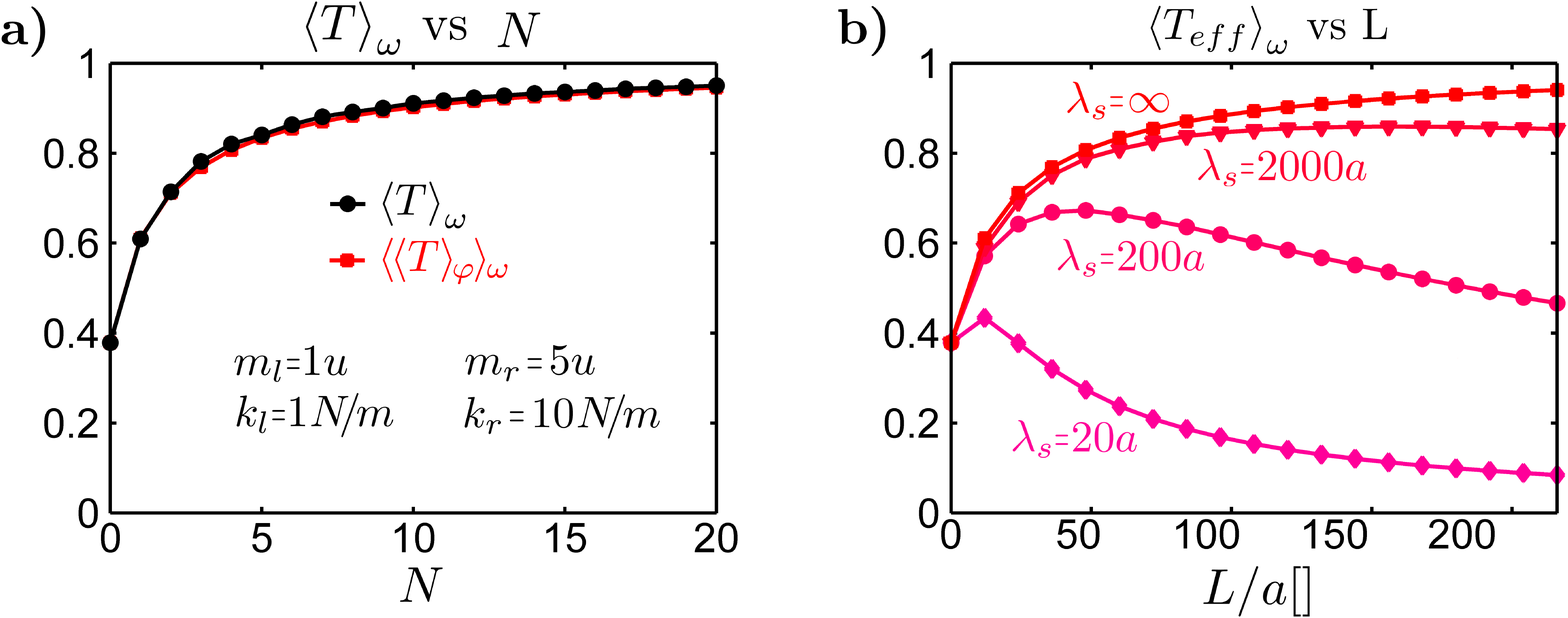}
	\caption{{\bf a)} Using the exponential variation of broadenings, as the number of junction materials $N$ increases, $\left\langle T\right\rangle_\omega$ and $\langle\langle T\rangle_\varphi\rangle_\omega$ approach to one. {\bf b)} As N increases the length between contacts increases and phonon flow is affected by scattering mechanism different from interfacial scattering. Therefore, there is a sweet spot for $N$ at which we stop increasing thermal conductance by adding junction materials.}
	\label{figAvgTvsL}
\end{figure}

Transmission when phonon phase is randomized or incoherent transmission $\langle T \rangle_{\varphi}$ can be found from the total resistance of the system similar to Eq.~\ref{equaddprop}  \cite{Anderson80,Datta97,Datta05}
\begin{equation}
\frac{1- \langle T \rangle_{\varphi}}{\langle T \rangle_{\varphi}}=\sum_j\frac{1-T_j}{T_j}
\label{equaddprop2}.
\end{equation}
Eq.~\ref{equaddprop2} becomes minimum {\it when all the factors on the right hand side are equal}. This condition needs the ratios of broadenings between consecutive materials to be the same and leads to the exponential variation in Eq.~\ref{equevGamma} (Fig.~\ref{figTNM}b). Replacing that variation into Eq.~\ref{equaddprop2} give us the maximum $\langle T \rangle_\varphi$, which can be approximated by 
\begin{equation}
\langle T \rangle_\varphi\approx \frac{4N}{\ln^2(\Gamma_l/\Gamma_r)+4N}
\end{equation}
when $N$ tends to infinite and by 
\begin{equation}
\langle T \rangle_\varphi\approx (T_{lr})^{(1/N+1)}
\end{equation}
for small $N$. As shown in Fig.~\ref{figAvgTvsL}a, $\langle T \rangle_\varphi$ increases as $N$ increases. This non intuitive result happens because the increment in resistance by the addition of an interface is dominated by the decrease of the interfacial resistance of all the interfaces due to less material mismatch. 

As the number of junction materials increases, the length between contacts also increases and other scattering mechanisms besides interfacial scattering become important. We can combine all of them using Matthiessen's rule to obtain an effective mean free path (mfp) in terms of the mfp for interfaces $\lambda_i$ and the mfp for other scattering mechanisms $\lambda_s$
\begin{equation}
\frac{1}{\lambda_{eff}}=\frac{1}{\lambda_{i}}+\frac{1}{\lambda_{s}},
\label{equlambdaeff}
\end{equation}
and an effective transmission
\begin{equation}
\langle T\rangle_\varphi^{eff} =\frac{\lambda_{eff}}{L+\lambda_{eff}}.
\label{equTeff}
\end{equation}
Note that Eq.~\ref{equlambdaeff} and \ref{equTeff} can be derived from equation \ref{equaddprop2} \cite{Datta97,Datta05}. We define the interfacial mfp for $N$ materials following the exponential variation $\lambda_i$ by replacing Eq.~\ref{equevGamma} into Eq.~\ref{equaddprop2}, defining the distance between interfaces as $L_i=L/N$ and comparing with Eq.~\ref{equTeff} so that
\begin{equation}
\langle T \rangle_\varphi=\frac{\lambda_i}{L+\lambda_i} \ \ \ \ \text{ and } \ \ \ \ \ \lambda_i=\left(\frac{LL_i}{L+L_i}\right)\frac{T_i}{R_i}.
\end{equation}
$\lambda_i$ is dominated by $L_i$ since we expect $L_i<L$ and is inversely proportional to the resistance at each interface. Thus, as $N$ increases $\lambda_i$ also increases. However, the effective mfp is bounded by $\lambda_s$. Thus, there is a sweet spot for the number of junction materials at which adding more interfaces does not increase the effective mfp and $\langle T\rangle_\varphi^{eff}$ is dominated by the increment of $L$ (Fig.~\ref{figAvgTvsL}b).

\section{Conclusion}

We show how to choose the properties of a single or multiple junction materials between prespecified contacts to maximize thermal conductance in 1D systems within the coherent and incoherent regimes. For a single junction our choice should be the GM of the contact broadenings, which requires two conditions: (1) the impedance equal to the GM of the contact impedances to favor low frequency phonons and (2) the cut-off frequency close to the minimum of the contact cut-off frequencies to favor high frequency phonons. Those sub conditions translate into rules to choose the junction mass, spring constant, group velocity, Debye temperature or density of states. These rules allow us to pinpoint the role of each quantity in bridging phonon across the interface. For multi-junctions materials we show that the GM condition evolves into an exponential variation of broadenings. This variation pushes thermal conductance to the maximum limit allowed by the contacts as the number of materials increases. We also integrate interfacial scattering with other scattering mechanisms. We find a sweet spot for the number of material caused by the interplay between smoothing the interface by impedance matching and enlarging the distance between contacts.

\appendix

\begin{acknowledgments}
C.A.P. and A.W.G. are grateful to Jingjie Zhang for useful discussions and to NSF for its support from NSF-CAREER (QMHP 1028883) and from NSF-IDR (CBET 1134311).
\end{acknowledgments}

\bibliography{bibpaper}% Produces the bibliography via BibTeX.

\end{document}